\documentstyle[sprocl,fleqn,epsf]{article}
 
\begin{document}

 \begin{center}
 SUPERGRAVITY UNIFIED MODELS \\
 \end{center}
 \bigskip
 \bigskip
 \begin{center}
 R. ARNOWITT$^a$ and PRAN NATH$^b$
 \end{center}
 \bigskip
 \bigskip
 \noindent
 $^a$ Center for Theoretical Physics, Department of Physics, Texas A\&M
 \mbox{University},\\
 \hspace*{.10in}College Station, TX  77843-4242\\
 \noindent
 $^b$ Department of Physics, Northeastern University, Boston, MA  02115
 \bigskip
 \begin{abstract}
 The development of supergravity unified models 
  and their implications
 for current and future experiment are discussed.
 \end{abstract}
 \bigskip
 \noindent
 {\bf 1.~~ Introduction}\\
 \medskip
 
 \indent
 Supersymmetry (SUSY) was initially introduced as a global symmetry \cite{golf} on
 purely theoretical grounds that nature should be symmetric between bosons and
 fermions.  It was soon discovered, however, that models of this type had a
 number of remarkable properties \cite{bzum}.  Thus the bose-fermi symmetry led to the
 cancellation of a number of the infinities of conventional field theories, in
 particular the quadratic divergences in the scalar Higgs sector of the Standard
 Model (SM).  Thus SUSY could resolve the gauge hierarchy problem that plagued
 the SM.  Further, the hierarchy problem associated with grand unified models
 (GUT) \cite{hgeo}, where without SUSY, loop corrections gave all particles GUT size
 masses \cite{egil} was also resolved.  In addition, SUSY GUT models with minimal
 particle spectrum raised the value for the scale of grand unification,
 $M_G$, to
 $M_G\cong 2\times 10^{16}$ GeV, so that the predicted proton decay rate 
 \cite{sdim,srab}
 was consistent with existing experimental bounds.  Thus in spite of the lack of
 any direct experimental evidence for the existence of SUSY particles,
 supersymmetry became a highly active field among particle theorists.
 
 However, by about 1980, it became apparent that global supersymmetry was
 unsatisfactory in that a phenomenologically acceptable picture of spontaneous
 breaking of supersymmetry did not exist.  Thus the success of the SUSY grand
 unification program discussed above was in a sense spurious in that the needed
 SUSY threshold $M_S$ (below which the SM held) could not be theoretically
 constructed.  In order to get a phenomenologically viable model, one needed
  ``soft breaking'' masses (i.e. supersymmetry breaking terms of dimenison
 $\leq$ 3 which maintain the gauge hierarchy \cite{lgir}) and these had to be
 introduced in an ad hoc fashion by hand.  In the minimal SUSY model, the
 MSSM \cite{fora},
 where the particle spectrum is just that of the supersymmeterized SM, one could
 introduce as many as 105 additional parameters (62 new masses and mixing angles
 and 43 new phases) leaving one with a theory with little predictive power.
 
 A resolution of the problem of how to break  supersymmetry spontaneously was
 achieved by promoting supersymmetry to a local symmetry i.e. supergravity \cite{dzfr}.
 Here gravity is included into the dynamics.  One can then construct
 supergravity [SUGRA] grand unified models \cite{chams,applied} where the spontaneous
 breaking of supersymmetry occurs in a ``hidden'' sector via supergravity
 interactions in a fashion that maintains the gauge hierarchy.  In such theories
 there remains, however, the question of at what scale does supersymmetry break,
 and what is the ``messenger'' that communicates this breaking from the hidden
 to the physical sector.  In this chapter we consider models where supersymmetry
 breaks at a scale $Q > M_G$ with gravity being the messenger \cite{alte}.  Such models
 are economical in that both the messenger field and the agency of supersymmetry
 breaking are supersymmetrized versions of  fields and interactions that already
 exist in nature (i.e. gravity).  In addition, models of this type may turn
 out to be consequences of string theory.
 
 The strongest direct evidence supporting supergravity GUT models is the
 apparent experimental grand unification of the three gauge coupling constants
 \cite{lang}.  This result is non-trivial not only because three lines do not
 ordinarily intersect at one point, but also because there is only a narrow
 acceptable window for $M_G$.  Thus one requires 
 $M_G{\tiny \begin{array}{l} >\\ \sim \end{array}} 5\times 10^{15}\ GeV$  
 so as not to violate current experimental bounds on proton decay
 for the $p\rightarrow e^+ +\pi^0$ channel (which occurs in almost all GUT
 models) and one requires $M_G\stackrel{<}{\sim} 5\times 10^{17}$ GeV $\cong
 M_{string}$ (the string scale) so that gravitational effects do not become
 large invalidating the analysis.   Further, assuming an MSSM type of particle
 spectrum between the electroweak scale $M_Z$ and $M_G$, acceptable grand
 unification occurs only with one pair of Higgs doublets and at most four
 generations.  Finally, naturalness requires that SUSY thresholds be at
 $M_S\stackrel{<}{\sim} 1$ TeV which turns out to be the case.  Thus the possibility
 of grand unification is tightly constrained.
 
 At present, grand unification in SUGRA GUTs can be obtained to within about 2-3
 std. \cite{chank,das}  
 However, the closeness of $M_G$ to the Planck scale,
 $M_{P\ell}$ = $(\hbar c/8\pi G_N)^{1/2}\cong 2.4 \times 10^{18}$ GeV,
 suggests the
 possibility that there are O($M_G/M_{P\ell}$) corrections to these models.  One
 might, in fact, expect such structures to arise in string theory as
 nonrenormalizable operators (NROs) obtained upon integrating out the tower of
 Planck mass states.  Such terms would produce $\approx$ 1\% corrections at
 $M_G$
 which might grow to $\approx$ 5\% corrections at $M_Z$.  Indeed, as will
 be seen
 in Sec.2, it is just such NRO terms involving the hidden sector fields that
 give rise to the soft breaking masses, and so it would not be surprising
 to find
 such structures in the physical sector as well.  Thus SUGRA GUTs should be
 viewed as an effective theory and, as will be discussed in Sec.4, with small
 deviations between theory and experiment perhaps opening a window to Planck
 scale physics.
 
 One of the fundamental aspects of the SM, not explained by that theory, is the
 origin of the spontaneous breaking of SU(2) x U(1).  SUGRA GUTS offers an
 explanation of this due to the existence of soft  SUSY 
 breaking masses at $M_G$.
 Thus as long as at least one of the soft breaking terms are present at $M_G$,
 breaking of SU(2) x U(1) can occur at a lower energy \cite{chams,inou},  
 providing
 a natural Higgs mechanism.  Further, radiative breaking occurs at the electroweak
 scale provided the top quark is heavy ie. 100 GeV
 $\stackrel{<}{\sim}m_t\stackrel{<}{\sim}$ 200 GeV.  The minimal SUGRA 
 model $^{10,17-19}$ (MSGM), which assumes universal soft
 breaking terms, requires only four
 additional parameters and one sign to describe all the interactions and masses
 of the 32 SUSY particles.  Thus the MSGM is a reasonably predictive model, and
 for that reason is the model used in much of the phenomenological analysis of
 the past decade.  However, we will see in Sec. 2 that there are reasons to
 consider non-universal extensions of the MSGM, and discuss some of the
 experimental consequences they produce in Sec.4.
 
 \bigskip
 \noindent
 {\bf 2.~~Soft Breaking Masses}\\
 
 \medskip
 \indent
 Supergravity interactions with chiral matter fields,
 $\lbrace\chi_i(x),\phi_i(x)\rbrace$ (where $\chi_i(x)$ are left (L) Weyl
 spinors and $\phi_i(x)$ are complex scalar fields) depend upon three functions of the
 scalar fields:  the superpotential $W(\phi_i)$, the gauge kinetic function
 $f_{\alpha\beta}(\phi_i,\phi_i^{\dag}$) (which enters in the Lagrangian as
 $f_{\alpha\beta}F^{\alpha}_{\mu\nu}F^{\mu\nu\beta}$ with $\alpha,\beta$ = gauge
 indices) and the Kahler potential $K(\phi_i,\phi_i^{\dag}$) (which appears in
 the scalar kinetic energy as
 $K_j^i\partial_{\mu}\phi_i\partial^{\mu}\phi_j^{\dag},$
 $K_j^i\equiv\partial^2 K^2 /\partial\phi_i\partial\phi_j^{\dag}$ and
 elsewhere).  W and K enter only in the combination
 \medskip
 \begin{equation}
 G(\phi_i,\phi_i^{\dag})=\kappa^2 K(\phi_i,\phi_i^+)+\ell n
 [\kappa^6\mid W(\phi_i)\mid^2]
 \end{equation}
 \medskip
 \noindent
 where $\kappa = 1/M_{P\ell}$.  Writing
 $\lbrace\phi_i\rbrace=\lbrace\phi_a, z\rbrace$ where $\phi_a$ are physical
 sector fields (squarks, sleptons, higgs) and z are the hidden sector fields
 whose VEVs $\langle z\rangle ={\cal O}(M_{P\ell})$ break supersymmetry, one
 assumes that the superpotential decomposes into a physical and a hidden part,
 \medskip
 \begin{equation}
 W(\phi_i) = W_{phy}(\phi_a, \kappa z) +W_{hid}(z)
 \end{equation}
 \medskip
 \noindent
 Supersymmetry breaking is scaled by requiring $\kappa^2 W_{hid} = {\cal
 O}(M_S)\tilde W_{hid}(\kappa z)$ and the gauge hierarchy is then
 guaranteed by the additive
 nature of the terms in Eq.(2).  Thus only gravitational interactions remain to
 transmit SUSY breaking from the hidden sector to the physical sector.
 
 A priori, the functions W, K and $f_{\alpha\beta}$ are arbitrary.
 However, they
 are greatly constrained by the conditions that the model correctly reduce to
 the SM at low energies, and that non-renormalizable corrections be scaled
 by $\kappa$
 (as would be expected if they were the low energy residue of string physics of
 the Planck scale).  Thus one can expand these functions in polynomials of the
 physical fields $\phi_a$
 
 \begin{equation}
 f_{\alpha\beta} (\phi_i,\phi_i^{\dag}) = c_{\alpha\beta} + \kappa
 d^a_{\alpha\beta}(x,y)\phi_a+\cdots
 \end{equation}
 \medskip
 \begin{equation}
 W_{phys}(\phi_i)=\frac{1}{6}\lambda^{abc}(x)\phi_a\phi_b\phi_c+\frac{1}{24}
 \kappa\lambda^{abcd}(x) \phi_a\phi_b\phi_c\phi_d+\cdots
 \end{equation}
 \medskip
 \begin{eqnarray}
 K(\phi_i,\phi_i^{\dag})&=& \kappa^{-2} c(x,y)+c^a_b(x,y)\phi_a\phi_b^{\dag}+
 (c^{ab}(x,y)\phi_a\phi_b + h.c.)\nonumber\\
 &+&\kappa(c^a_{bc}\phi_a\phi_b^{\dag}\phi_c^{\dag} + h.c.)+\cdots
 \end{eqnarray}
 \medskip
 \noindent
 x=$\kappa z$ and y =$\kappa z^{\dag}$, so that $\langle x\rangle$,
 $\langle y\rangle = {\cal O}$ (1).  The scaling hypothesis for the NRO's imply
 then that the VEVs of the coefficients $c_{\alpha\beta},~c_{\alpha\beta}^a,
 ~\lambda^{abc},~c,~c_b^a,~c^{ab}$ etc. are all ${\cal O}$(1).
 
 The holomorphic terms in K labeled by $c^{ab}$ can be transformed from K to W
 by a Kahler transformation, $K\rightarrow K-(c^{ab}\phi_a\phi_b + h.c.)$ and
 \medskip
 
 \begin{equation}
 W\rightarrow W exp [\kappa^2  c^{ab}~\phi_a\phi_b] = W +
 {\tilde\mu}^{ab}\phi_a\phi_b + \cdots
 \end{equation}
 \medskip
 \noindent
 where $\tilde\mu^{ab}(x,y)=\kappa^2 Wc^{ab}$.  Hence  $\langle
 \tilde{\mu}^{ab}\rangle=\cal O (M_S)$, and one obtains a $\mu$-term with the 
 right order of magnitude after SUSY breaking provided only that $c^{ab}$ is not 
 zero \cite{soni}.
 The cubic terms in W are just the Yukawa couplings with
 $\langle\lambda^{abc}(x)\rangle$ being the Yukawa coupling constants.  Also
 $\langle c_{\alpha\beta}\rangle =\delta_{\alpha\beta}$, $\langle
 c_b^a\rangle=\delta_b^a$ and $\langle c_{xy}\rangle$ = 1 ($c_x\equiv\partial
 c/\partial x$ etc.) so that the field kinetic energies have canonical
 normalization.
 
 The breaking of SUSY in the hiddden sector leads to the generation of a series
 of soft breaking terms $^{10,11,17-21}$. 
 We consider here the case where $\langle
 x\rangle = \langle y\rangle$ (i.e. the hidden sector SUSY breaking does not
 generate any CP violation) and state the leading terms.  Gauginos gain a soft
 breaking mass term at $M_G$ of
 ($m_{1/2})_{\alpha\beta}\lambda^{\alpha}\gamma^0\lambda^{\beta}$
 ($\lambda^{\alpha}$ = gaugino Majorana field) where
 
 \begin{equation}
 (m_{1/2})_{\alpha\beta}= \kappa^{-2}\langle
 G^i (K^{-1})^i_j Re f_{\alpha\beta j}^{\dag}\rangle m_{3/2}
 \end{equation}
 
 \noindent
 Here $G^i\equiv\partial G/\partial\phi_i$, $(K^{-1})^i_j$ is the matrix
 universe of $K_j^i$, $f_{\alpha\beta j} =\partial
 f_{\alpha\beta}/\partial\phi_j^{\dagger}$ and $m_{3/2}$ is the gravitino mass:
 $m_{3/2} =
 \kappa^{-1}\langle exp [G/2]\rangle$.  In terms of the expansion of Eqs.(3-5)
 one finds

 \begin{equation}
 (m_{1/2})_{\alpha\beta}= [c+\ell n (W_{hid})]_x Re~c_{\alpha\beta y}^{\ast}
 m_{3/2}
 \end{equation}
 
 \noindent
 and $m_{3/2}$ = (exp $\frac{1}{2} c)\kappa^2 W_{hid}$ where it is
 understood from
 now on that x is to be replaced by its VEV in all functions (e.g.
 c(x)$\rightarrow c(\langle x\rangle) = {\cal O}(1))$ so that 
 $m_{3/2} = {\cal O}(M_S).$  
 One notes the following about Eq.(8):  (i)  For a
 simple GUT group, gauge invariance implies that $c_{\alpha\beta}\sim
 \delta_{\alpha\beta}$ and so gaugino masses are universal at mass scales above
 $M_G$.  (ii)  While $m_{1/2}$ is scaled by $m_{3/2} = {\cal O}(M_S)$, it can
 differ from it by a significant amount.  
 (iii)  From Eq.(8) one sees that it is
 the NRO such as $\kappa zm_{3/2}\lambda^{\alpha}\gamma^0\lambda^{\alpha}$ that
 gives rise to $m_{1/2}$.  Below $M_G$, where the GUT group is broken, the
 second term of Eq.(3) would also contribute yielding a NRO of size $\kappa
 d^a_{\alpha\beta}\phi_a m_{3/2}$ $\lambda^{\alpha}\gamma^0\lambda^{\beta}\sim
 (M_G/M_{P\ell})$ $m_{3/2} \lambda\gamma^0\lambda$ \cite{hill} for fields
 with VEV
 $\langle \phi_a \rangle = {\cal O} (M_G)$ which break the GUT group.  Such
 terms give small corrections to the universality of the gaugino masses and
 effect grand unification.  They are discussed in Sec.4.
 
 The effective potential for the scalar components of chiral multiplets is given
 by \cite{chams,crem}

 $$
 V= exp [\kappa^2 K]~[( K^{-1})^j_i( W^i+\kappa^2 K^i
 W)~(W^j+\kappa^2 K^j W)^{\dag}
 $$
 \begin{equation}
 \quad \quad \quad \quad \quad \quad  
 \quad \quad \quad 
-3\kappa^2\mid W\mid^2] + V_D
 \end{equation}
 \medskip
 \noindent
 where $W^i=\partial W/\partial\phi_i$ etc., and 
 
 \begin{equation}
 V_D = \frac{1}{2}
 g_{\alpha}g_{\beta}~(Ref^{-1})_{\alpha\beta}~(K^i(T^{\alpha})_{ij}\phi_j)~(
 K^k(T^{\beta})
 _{k\ell}\phi_{\ell})
 \end{equation}
 
 \noindent
 where $g_{\alpha}$ are the gauge coupling constants.  Eqs.(2-6) then lead to
 the following soft breaking terms at $M_G$:
 
 \begin{equation}
 V_{soft} = ( m_0^2)^a_b~\phi_a\phi_b^{\dag}+\biggl[\frac{1}{3}
 {\tilde A}^{abc}\phi_a\phi_b\phi_c + \frac{1}{2}{\tilde
 B}^{ab}\phi_a\phi_b+h.c.\biggr]
 \end{equation}
 
 \noindent
 In the following, we impose the condition that the cosmological constant vanish
 after SUSY breaking, i.e. $\langle V\rangle = 0$.  [This is a fine tuning of
 ${\cal O} (M_S^2M_{P\ell}^2)$.]   
 Then one finds for D-flat breaking 
 
 \begin{equation}
 (m_0^2)^a_b =[3 (c_{cx}^a~c_{by}^c-c^a_{b~xy})+\delta_b^a]m_{3/2}^2
 \end{equation}
 
 \begin{equation}
 {\tilde A}^{abc} = \pm {\sqrt 3} \biggl[
 \biggl\{ c_x h^{abc} - (c_{dx}^ah^{dac} + c_{dx}^b h^{dac}+ c_{dx}^c  h^{abd})
 \biggr\}
 + h^{abc}_x\biggr] m_{3/2}
 \end{equation}
 
 \begin{equation}
 {\tilde B}^{ab}=\biggl[\biggl\{(\pm{\sqrt 3} c_x -1)\mu^{ab}\mp{\sqrt 3}
 (c_{dx}^a\mu^{db}+c^b_{dx}\mu^{ad})\biggr\}\pm{\sqrt 3} \mu^{ab}_x\biggr]~m_{3/2}
 \end{equation}
 
 \noindent
 where the sign ambiguity arises from the two roots of the equation $\langle V
 \rangle = 0$.  In Eqs.(12-14), $h^{abc}=\langle exp
 (\frac{1}{2}c)\lambda^{abc}\rangle$ are the Yukawa coupling constants and
 $\mu^{ab}=\langle exp (\frac{1}{2}c)~{\tilde\mu}^{ab}\rangle$ is the
 $\mu$-parameter, $h_x^{abc}\equiv \langle exp
 (\frac{1}{2}c)~\partial\lambda^{abc}/\partial x\rangle$ and
 $\mu_x^{ab}\equiv \langle exp (\frac{1}{2}
 c)\partial{\tilde\mu}^{ab}/\partial x\rangle$.  Since K is hermitian,
 $c_b^a$ (defined in Eq.(5)) obeys $c_b^a = c_a^{b*}$.
 
 One notes the following about Eqs.(12-14):  (i)  The scalar soft breaking
 masses ($m_0^2)^a_b$ are in general not universal unless the Kahler metric is
 flat, i.e. $c_b^a=\delta_b^a$ \cite{chams,soni}, or more generally unless the fields z
 which break SUSY couple universally to the physical sector i.e. $c_b^a ={\tilde
 c}(x,y)\delta_b^a$.  (ii)  $\tilde A^{abc}$ can contain a term,
 $\lambda_x^{abc}$ which
 is not scaled by the Yukawa coupling constants $h^{abc}$ if
 $\lambda^{abc}$ is a function of the z fields.  
 This possibility may occur in string theory,
 where the role of z is possibly played by the dilaton and moduli 
 fields \cite{kapl}.
 In the following we will neglect the $\lambda_x^{abc}$ term 
 (and similarly the $\mu_x^{ab}$ term in Eq.(14)).
 
 The soft breaking masses in Eqs.(12-14) depend upon two hermitian matrices
 $c^a_{bx}$ and $c^a_{bxy}$ and the parameter $c_x$.  The number of new SUSY
 parameters this implies depends upon the GUT group.  We consider here the
 example of SU(5) supergravity GUT with R parity invariance.  The superpotential
 is given by
 
 \begin{eqnarray}
 W_{phys} &=& h_{ij}^{(1)}M_i^{XY} {\overline M}_{Xj}H_{1Y}+\frac{1}{4}
 h^{(2)}_{ij} \epsilon_{XYZWU} M_i^{XY}M_j^{ZW}H_2^U
 \nonumber\\
 &+&\mu H_2^X{\overline H}_{1X} +
 W_{GUT}
 \end{eqnarray}
 \medskip
 \noindent
 where X,Y,$\cdots$ = 1$\cdots$ 5 are SU(5) indices, i,j = 1,2,3 are generation
 indices, $M_i^{XY}$ and ${\overline M}_{Xi}$ are 10 and ${\overline 5}$ matter
 fields and $H_2^X$ and ${\overline H}_{1X}$ are 5 and ${\overline 5}$ Higgs
 fields.  $W_{GUT}$ contains additional Higgs fields which break SU(5) at
 $M_G$.  $H_2$ and $\overline H_1$ contain the pair of light Higgs doublets
 below $M_G$.  The soft breaking parameters then lead to 42 additional
 parameters for this model (25 new masses and mixing angles and 17 new 
 phases),  considerably fewer than the 105
 additional parameters of the MSSM.  
 While this is still a formidable number, 
 many are experimentally uninteresting (e.g. scaled by the small
 Yukawa coupling constants), and we will see in Sec.4 that the remaining
 parameters are few enough (at
 least for the CP conserving sectors of the theory) to obtain interesting
 experimental predictions.
 
 \bigskip
 \noindent
 {\bf 3.~~Radiative Breaking and the Low Energy Theory}\\
 \medskip
 \indent
 
 In Sec.2, the SUGRA GUT model above the GUT scale i.e. at $Q >M_G$ was
 discussed.  Below $M_G$ the GUT
 group is spontaneously broken, and we will assume here that the SM group,
 SU(3) x SU(2) xU(1), holds for $Q < M_G$ \cite{so(10)}.  
 Contact with accelerator
 physics at low
 energy can then be achieved using the renormalization group equations (RGE)
 running from $M_G$ to the electroweak scale $M_Z$.  As one proceeds
 downward from
 $M_G$, the coupling constants and masses evolve, and provided at least one soft
 breaking parameter and also the $\mu$ parameter at $M_G$ is not zero, the
 large top
 quark Yukawa can turn the $H_2$ running (mass)$^2$, $m_{H_{2}}^2(Q)$,
 negative at
 the electroweak scale \cite{inou}. Thus the spontaneous breaking of
 supersymmetry at
 $M_G$ triggers the spontaneous breaking of SU(2) xU(1) at the electroweak
 scale.
 In this fashion all the masses and coupling constants at the electroweak
 scale can
 be determined in terms of the fundamental parameters (Yukawa coupling constants
 and soft breaking parameters) at the GUT scale, and the theory can be
 subjected to
 experimental tests.
 
 The conditions for electroweak symmetry breaking arise from minimizing the
 effective potential V at the electroweak scale with respect to the Higgs VEVs
 $v_{1,2} = \langle H_{1,2}\rangle$.  This leads to the equations \cite{inou}
 
 \begin{equation}
 \mu^2 =\frac{\mu_1^2-\mu_2^2tan^2\beta}{tan^2\beta-1} - \frac{1}{2}
 M_Z^2;~~~sin^2\beta = \frac{-2B\mu}{2\mu^2+\mu_1^2+\mu^2_2}
 \end{equation}
 
 \noindent
 where $tan\beta = v_2/v_1$, B is the quadratic soft breaking parameter
 ($V_{soft}^B=B\mu H_1H_2$),  $\mu_i=m_{H_{i}}^2+\Sigma_i$, and
 $\Sigma_i$ are loop corrections \cite{gamb}.
 All parameters are running parameters at the
 electroweak scale which we take for convenience to be $Q=M_Z$ .  Eq.(16) then
 determines the $\mu$ parameter and allows the elimination of B in terms of
 $tan\beta$.  This determination of $\mu$ greatly enhances the predictive
 power of the model.
 
 We parameterize the Higgs soft breaking masses at $M_G$ by
 
 \begin{equation}
 m_{H_{1}}^2 = m_0^2 (1+\delta_1);~~m_{H_{2}}^2=m_0^2 (1+\delta_2)
 \end{equation}
 
 \noindent
 and the third generation of sfermion masses by
 
 \begin{eqnarray}
 m_{q_{L}}^2&=&m_0^2(1+\delta_3);~~m_{u_{R}}^2=m_0^2(1+\delta_4);~~m_{e_{R}}
 ^2=m_0^2(1+\delta_5)\nonumber\\
 m_{d_{R}}^2&=&m_0^2
 (1+\delta_6);~~m_{\ell_{L}}^2=m_0^2(1+\delta_7);
 \end{eqnarray}
 
 \noindent
 where $q_L=(u_L,~d_L)$, $\ell_L =(\nu_L,~e_L)$ etc.  The $\delta_i$ thus
 represent non-universal deviations from the reference mass $m_0$.  For GUT
 groups
 which contain an SU(5) subgroup (e.g. SU(N), N$\geq$5; SO(N), N$\geq$10, $E_6$
 etc.) with matter embedded in the 10 and $\overline 5$ representations, one has
 
 \begin{equation}
 \delta_3=\delta_4=\delta_5;~~~\delta_6=\delta_7
 \end{equation}

 In the following we will choose $m_0$ to be the soft breaking mass of the
 first two
 generations (assumed universal to suppress FCNC effects \cite{sdim}).  
 We will also keep
 only the large top Yukawa coupling constant $h_t$ and its associated 
 cubic soft SUSY 
 breaking parameter $A_t$ (an approximation valid for
 $tan\beta\stackrel{<}{\sim}$ 20).  
 This allows an analytic solution of the 1-loop RGE and enables one to
 understand physically the predictions of the model.
 One finds \cite{nonuni}
 
 \begin{eqnarray}
 \mu^2&=&{t^2\over t^2-1}\biggl[\biggl({1-3D_0\over 2}
 +{1\over t^2}\biggr)
 +\biggl({1-D_0\over 2}(\delta_3+\delta_4)
 -{1+D_0\over 2}\delta_2\nonumber\\
 &+&{1\over t^2} \delta_1\biggr)\biggr] m_0^2
 +{t^2\over t^2-1}\biggl[{1\over 2}(1-D_0){A_R^2\over D_0}
 +C_{\mu}m_{1/2}^2\biggr]
 -{1\over 2} M_Z^2\nonumber\\
 &+&{1\over 22} {t^2+1\over t^2-1} S_0\biggl(
 1-{\alpha_1(Q)\over\alpha_G}\biggr)
 \end{eqnarray}
 
 \noindent
 where $C_{\mu} = {1\over 2} D_0(1-D_0)~(H_3/F)^2 + e-g/t^2$, $t\equiv
 tan\beta$, and $D_0\cong 1-m_t^2/(200 sin\beta)^2$.  
 $D_0$ vanishes at the t-quark Landau pole
 (for $m_t$ = 175 GeV, $D_0\leq$ 0.23) and $A_R=A_t-m_{1/2} (H_2-H_3/F)$ is
 the residue at the Landau pole 
 ($
 A_R \cong A_t - 0.61 (\alpha_3/\alpha_G) m_{1/2}
 $),
 i.e. $A_0 = A_R/D_0-(H_3/F)m_{1/2}$ where $A_0$ is the cubic
 soft SUSY breaking parameter at $M_G$ \cite{nwa}.  $\alpha_G$ is the GUT
 coupling constant
 $(\alpha_G\simeq 1/24$) and
 
 \begin{equation}
 S_0=Tr(Ym^2)=(m_{H_{2}}^2-m_{H_{1}}^2) +\sum^{ng}_{i=1} (m_{q_{L}}^2 - 2
 m_{u_{R}}^2+m_{d_{R}}^2-m_{\ell_{L}}^2 +m_{e_{R}}^2)
 \end{equation}
 
 \noindent
 In Eq.(20), the form factors e.g. $H_2$, $H_3$, $F$ are given in
 Iba\~nez et
 al \cite{inou}, $n_g$ is the number of generations, and the
 (mass)$^2$ in Eq.(21) are all
 to be taken at $M_G$.  
 We note that $S_0$ vanishes for universal soft breaking
 masses, and reduces to $S_0=m_{H_{2}}^2-m_{H_{1}}^2$ for three generation GUT
 models obeying Eq.(19).  (The coefficient (1-$\alpha_1
 (M_Z)/\alpha_G)/22\cong$
 0.0268 and so this term is generally small.)
 
 The renormalization group equations evolve the universal gaugino mass
 $m_{1/2}$ at
 $M_G$ to separate masses for SU(3), SU(2) and U(1) at $M_Z$:
 
 \begin{equation}
 \tilde m_i =(\alpha_i(M_Z)/\alpha_G)m_{1/2};~~~i = 1,2,3
 \end{equation}
 
 \noindent
 where at 1-loop, the gluino mass $m_{\tilde g}=\tilde m_3$ \cite{mart}.
  The breaking of
 SU(2) x U(1) causes a mixing of the gaugino-higgsino states giving rise to two
 charginos $\chi_i^{\pm}$, {\it i}=1,2 and four neutralinos $\chi_i^0$,
 {\it i} =
 1$\cdots$ 4.  In the $\tilde W^{\pm}-{\tilde H}^{\pm}$ basis, the chargino mass
 matrix is
 \medskip
 \begin{equation}
 M_{\chi}^{\pm}=\left(\begin{array}{cc}
 {\tilde m}_2 & \sqrt 2 M_W sin\beta\\
 \\
 \sqrt 2M_W cos\beta &\mu
 \end{array}\right)
 \end{equation}
 \medskip
 
 \noindent
 and in the ($\tilde W_3,~\tilde B,~\tilde H_1,~\tilde H_2$) basis and the
 neutralino mass matrix is
 \medskip
 \begin{equation}
 M_{\chi}^0=\left(\begin{array}{cccc}
 \tilde m_2 & o & a & b\\
 \\
 o & \tilde m_1 & c & d\\
 \\
 a & c & o & -\mu\\
 \\
 b & d & -\mu & o
 \end{array}\right)
 \end{equation}
 
 \medskip
 \noindent
 where $a=M_Z cos\theta_W cos\beta$, $c=-tan\theta_Wa$, $b=-M_Z
 cos\theta_W$ sin$\beta$,  d=-tan$\theta_Wb$ and $\theta_W$ is the weak
 mixing angle.  In the domain where $\mu^2/M_Z^2 >>  1$ one has
 
 \begin{eqnarray}
 2m_{\chi_{1}}^0&\cong& m_{\chi_{1}}^{\pm}\cong m_{\chi_{2}}^0\simeq {1\over 3}
 m_{\tilde g}\nonumber\\
 m_{\chi_{3}}^0 &\cong& m_{\chi_{4}}^0 \cong m_{\chi_{2}}^{\pm}\simeq \mu >>  
 m_{\chi_{1}}^0
 \end{eqnarray}
 \medskip
 \noindent
 i.e. the light states are mostly gauginos and the heavy states mostly
 higgsinos.
 
 The large top mass also causes L-R mixing in the stop mass matrix
 
 \medskip
 \begin{equation}
 M_{\tilde t}^2 =\left(\begin{array}{cc}
 m_{t_{L}}^2 & -m_t(A_t+\mu ctn\beta)\\
 \\
 -m_t(A_t+\mu ctn\beta) & m_{t_{R}}^2
 \end{array}\right)
 \end{equation}  
 \medskip
 \noindent
 where
 
 \begin{equation}
 m_{t_{L}}^2 = m_{Q_{L}}^2 + m_t^2 + \left({1\over 2}-{2\over 3} sin^2
 \theta_W\right) M_Z^2 cos 2\beta
 \end{equation}
 
 \begin{equation}
 m_{t_{R}}^2 = m_U^2 + m_t^2 + {2\over 3} sin^2\theta_W M_Z^2 cos 2\beta
 \end{equation}
 
 \noindent
 and
 
 \begin{eqnarray}
 m_{Q_{L}}^2&=&\left[\left({1+D_0\over2}\right) + {5+D_0\over 6}\delta_3 -
 {1-D_o\over 6} (\delta_2+\delta_4)\right] m_0^2\nonumber\\
 &-& {1\over 6}(1-D_0) {A_R^2\over D_0} + C_Q m_{1/2}^2 - {1\over 66} S_0
 (1-\alpha_1(Q)/\alpha_G)
 \end{eqnarray}
 
 \begin{eqnarray}
 m_U^2&=& \left[D_0 + {2+D_0\over 3} \delta_4 -
 {1-D_0\over
 6} (\delta_2+\delta_3)\right] m_0^2\nonumber\\
 &-& {1\over 3} (1-D_0) {A_R^2\over D_0} + C_U m_{1/2}^2 + (2/33) S_0
 (1-\alpha_1(Q)/\alpha_G)
 \end{eqnarray}
 
 \noindent
 The coefficients $C_Q$, $C_U$ are given by
 
 \begin{equation}
 C_Q = {1\over 3} \left[-{1\over 2} D_0(1-D_0) (H_3/F)^2+e\right]+{\alpha_G\over
 4\pi} \left({8\over 3} f_3 + f_2-{1\over 15} f_1\right)
 \end{equation}
 
 \begin{equation}
 C_U ={2\over 3}\left[ - {1\over 2} D_0(1-D_0) (H_3/F)^2+e\right] +
 {\alpha_G\over
 4\pi} \left( {8\over 3} f_3-f_2+{1\over 3} f_1\right)
 \end{equation}
 
 \noindent
 the form factors $f_i$ being given in Iba$\tilde n$ez et al. \cite{inou}.
 
 The simplest model is the one with universal soft breaking masses, i.e.
 $\delta_i$=0.  This model depends on only the four SUSY parameters and one
 sign
 \medskip
 \begin{equation}
 m_0, \quad m_{1/2}, \quad A_0, \quad B_0, \quad sign(\mu)
 \end{equation}
\medskip
 \noindent
 Alternately one may choose
 $m_0\  $,  $m_{\tilde g}\  $, $A_t\  $, $tan\beta\  $, and sign($\mu$)
  as the independent parameters.
 This case has been extensively discussed in the literature \cite{ross}.
 As can be seen above, however, the deviations from universality can
 significantly
 effect the values of $\mu^2$ and the stop masses, and these parameters play a
 crucial role in predictions of the theory.  Thus more recently, efforts to
 include
 non-universalities into calculations have been made, and some of these effects
 will be discussed in the context of dark matter in Sec.4.

\bigskip
\noindent
 {\bf 4.~~ Phenomenological Implications of Supergravity Models}\\
\medskip

\indent
 We shall discuss the implications of supergravity models mostly
 in the context of the parameter space of  Eq.(33) although many of the
 results of the analysis are valid for a broader class of supergravity
 models which include non-universalities as discussed in Secs.2 and 3.
 We shall analyse the effect of non-universalities  discussed in Sec.3 in some
 detail in the context of dark matter detectors.
 There are many phenomenological implications of supergravity unified
 models. One of the important implications concerns the signatures of
 supersymmetry in collider experiments. Such signatures depend crucially
 on the stability of the lowest supersymmetric particle (LSP).
 A stable LSP can be achieved by the imposition of an R parity symmetry.
 This is an attractive possibility as it also provides a candidate
 for cold dark matter. In fact using the renormalization group analysis
 discussed earlier, one finds
 that over most of the parameter space of
  supergravity unified models the lightest neutralino is the
  LSP \cite{lsp}.
  In this circumstance the decay of the
  supersymmetric particles will have missing energy signals  and
   at least  one of the carriers of missing energy will be the
  neutralino. Signals of this type were studied early on after the
  advent of supergravity models in the supersymmetric decays of the
  W and Z bosons \cite{swein} and such analyses have since been extended
  to the decays of all of the supersymmetric particles. Using these
  decay patterns one finds  a variety of  supersymmetric signals for SUSY
  particles at colliders where SUSY particles are expected to be
  pair produced when sufficient energies in the center of mass system
  are achieved.  One signal of special interest in the search for
  supersymmetry is the trileptonic signal \cite{trilep}.
  For example in $p\bar p$
  collisions one can have $p\bar p\rightarrow \tilde \chi^{\pm}_1+
  \tilde \chi^0_2+X
  \rightarrow(l_1\bar \nu_1\tilde \chi^0_1)+(l_2\bar l_2\tilde \chi^0_1)+X$ which
  gives a signal of three leptons and missing energy. Using this signal
  a chargino of mass up to 230 GeV can be explored at the Tevatron with
  an integrated luminosity of \cite{amidei}  10$fb^{-1}$.
 
         One of the most accurate probes of new physics beyond the
         Standard Model (SM) is provided by (g-2) for the muon and for the
         electron \cite{kino}.
         Currently the experimental value of $a_{\mu} \equiv (g_{\mu}/2
         -1)$ is $a_{\mu}^{exp}= 11659230(84) \times 10^{-10}$
         where as  $a_{\mu}^{theory}(SM)= 11659172(15.4) \times
         10^{-10}$. Here $a_{\mu}^{theory}(SM)$ contains the $O(\alpha^5)$
         Q.E.D. contributions, $O(\alpha^2)$ hadronic corrections
         and up to two loop corrections from the SM electro-weak sector.
         Essentially all of the error in the theoretical computations comes
         from the
         hadronic contributions which give
         $a_{\mu}^{hadron}(SM)= 687.0(15.4) \times 10^{-10}$ while the
         contribution from the electro-weak sector  
         has a very small error \cite{czar}, i.e.,
         $a_{\mu}^{EW}(SM)= 15.1(0.4) \times 10^{-10}$. It is expected
         that the new experiment underway at Brookhaven will reduce the
         experimental uncertainty by a factor of about 20. Further, a
         new analysis by Alemany et.al. \cite{alemany}
         using the $\tau$ data at LEP indicates a significant
         reduction in the hadronic error, i.e.,
         $a_{\mu}^{hadron}(SM)= 701.4(9.4) \times 10^{-10}$. The hadronic
         error  is expected to reduce further perhaps by another factor of
         two when new data from VEPP-2M, DA$\Phi$NE, and BEPC experiments
         comes in. Thus the improved determination of $g_{\mu}-2$
         in the Brookhaven experiment as
         well as the expected  reduction in the hadronic error
         will allow one to test the electro-weak SM contribution.
 
          It was pointed out in ref. \cite{yuan}
          that any experiment which tests the
         SM electro-weak contribution to $g_{\mu}-2$
         can also test the supersymmetric
         contribution. This conclusion is supported by the recent precision
         analyses of $g_{\mu}-2$ using the renormalization
         group \cite{lopez,chatto}. Thus the
         Brookhaven experiment is an important  probe
         of supergravity grand unification.
 
         Flavor changing neutral current processes also provide
         an important constraint on supergravity unified models.
         A process of great interest here is the decay $b\rightarrow
         s+\gamma$ which has been observed by the CLEO
         collaboration \cite{alam}
         with a branching ratio of BR($b\rightarrow  s+\gamma$)
         =(2.32$\pm 0.67$)$\times 10^{-4}$. In the SM this  decay proceeds
         at the loop level  with the exchange of W and Z bosons and the
         most recent analyses \cite{buras}
          of the branching ratio including leading and
         most of the next to  leading order QCD corrections yield
           BR($b\rightarrow  s+\gamma$) =(3.48$\pm 0.31$)$\times 10^{-4}$
           for $m_t=176$ GeV. In supersymmetry there are additional diagrams
           which contribute to this process \cite{bert}.
           Thus in SUGRA unification
           one has contributions from the exchange of  the charged Higgs,
           the charginos, the neutralinos and from the gluino.
           Interestingly one finds
           that while  the contribution from the charged Higgs exchange
           is always positive \cite{hewett} over much of the parameter
           space its effects are small since $H^{\pm}$ is heavy.  
   The contribution 
   from the
   exchange of the other SUSY particles  can be either positive or 
   negative with the
           contribution of the charginos being usually the dominant
           one \cite{diaz}.
           Thus supergravity can accomodate a value  of the
           BR($b\rightarrow  s+\gamma$) which is lower than the
           SM value \cite{diaz,na}.
           If the more accurate experimental determination of
            BR($b\rightarrow  s+\gamma$) continues to show a trend
            with a value of BR($b\rightarrow  s+\gamma$)
           lower than what the SM predicts, the results
            would provide a strong hint for a low lying chargino and a low
            lying third generation squark. The $b\rightarrow  s+\gamma$
          experiment also imposes an important constraint on dark
          matter analyses and we discuss this below.

    As mentioned earlier one of the remarkable results of
    supergravity grand unification with R parity invariance is the
    prediction that  the  lightest neutralino $\chi_1^0$ is the LSP over 
    most of the
    parameter space.  In this part of the
    parameter space  
     the $\chi_1^0$ is a
    candidate for cold dark matter (CDM).
     We discuss now the relic density of $\chi_1$ within the framework
     of the Big Bang Cosmology. The quantity that is computed theoretically
     is $\Omega_{\chi_1} h^2$ where  $\Omega_{\chi_1}=\rho_{\chi_1}/\rho_c$,
      $\rho_{\chi_1}$ is the neutralino relic density and $\rho_c$
     is the critical relic density needed to close the universe,
    $\rho_c\ =\ 3H^{2 }/8\pi G_N$, and $H=h 100 km/s Mpc$ 
   is the Hubble 
    constant. The  most conservative constraint on  $\Omega_{\chi_1} h^2$ is
     $\Omega_{\chi_1} h^2 < 1$. A variety of more constraining possibilities
     have also been discussed in the literature.
         One of the important elements in the computation of the relic
         density concerns the correct thermal averaging of the quantity
         ($\sigma v$) where  $\sigma$ is the neutralino  annihilation
         cross section in the early universe and 
    $v$ is the relative neutralino velocity.     Normally
         the thermal average is calculated by first making the approximation
         $\sigma v=a+b v^2$ and then evaluating its thermal
         average \cite{kolb}.
         However, it is known that such an approximation breaks down
         in the vicinity of thresholds and poles \cite{greist}.
         Precisely such a
         situation exists for the case of the annihilation of the
         neutralino through the Z and Higgs poles. An accurate analysis
         of the neutralino relic density in the presence of Z and Higgs
         poles was given in ref. \cite{dark1} and  similar analyses  have
         also been carried out since  by other authors \cite{bb,heidel}.
 
           There are a number of possibilites for  the  detection of
           dark matter both direct and indirect \cite{jungman,heidel}.
           We discuss here the
           direct method which involves the scattering of incident neutralino
           dark matter in the Milky Way from nuclei in terrestial targets. 
    The event
           rates
        consist of two parts \cite{goodman}: one involves an axial interaction
           and the other a scalar interaction. The axial
           (spin dependent) part $R_{SD}$
           falls off as $R_{SD}\sim 1/M_N$ for large $M_N$ where
           $M_N$ is the mass of the target nucleus, while
           the scalar (spin independent) part behaves as
           $R_{SI}\sim M_N$ and increases with $M_N$. Thus for heavy
           target nuclei the spin
           independent part $R_{SI}$ dominates over most of the
           parameter space of the model. Analysis of event rates
           for the scattering  of neutralinos off targets such as
           Ge and Xe indicates that the event rates can lie over a
           wide range, from O(1) event/kg d to O$(10^{-(5-6)})$
           event/kg d \cite{dark2}.
            Inclusion of non-universalities is seen to
           produce definite signatures in the event rate
           analysis \cite{nonuni,berez}.
           Thus the minimum event rates in the region $m_{\chi_1}< 65$
           GeV can be enhanced (for the case $\delta_1=-1=-\delta_2$)
           or reduced (for the case $\delta_1=1=-\delta_2$) by
           a factor of O(10-100) due to the presence of non-universalities  
        \cite{nonuni},
           while the effect of non-universalities is generally small for 
           $m_{\chi_{1}}>65\ GeV$.
         To illustrate how this  comes about we see that from Eq.(20) the
         non-universality correction to $\mu$ is given by
\begin{equation}
 \Delta\mu^2\ =\ {t^2\over t^2-1}\biggl[
 \biggl({1-D_0\over 2}(\delta_3+\delta_4)
 -{1+D_0\over 2}\delta_2
  + {1\over t^2} \delta_1\biggr)\biggr] m_0^2
 \end{equation}
    $\mu^2$ influences detection event rates in that an increase (decrease)
    in $\mu^2$ reduces (increases) $R_{SI}$.    
         Thus for $\delta_1=-1=-\delta_2$ one finds that
         $\Delta\mu^2$ is negative and tends to raise $R_{SI}$. 
         In addition,
         this could drive $\mu^{2}$ negative, 
          eliminating such points from the parameter space of the theory.
          These effects are more significant for small tan$\beta$ values 
         and thus drive the  minimum
          event rates higher in the region $m_{\chi_1}<65$ GeV. 
     For $\delta_1=-1=-\delta_2$ one has    $\Delta\mu^2>0$ and the
           effect here is opposite to that for the previous case.
           (Note also since $D_0$ is small, $\delta_3$ and $\delta_4$ can
          cancel or enhance the effects of $\delta_1$ and $\delta_2$). 
           Further, it is found that the event rates for  $\mu<0$
             are much suppressed (by a factor $\approx 100$)
            compared to the case $\mu>0$ \cite{dark2}. 
            This feature
            persists in the presence of non-\break universalities  
        \cite{nonuni}.

             The detection sensitivity of current detectors is typically
            of order of a few events/kg d \cite{bernabei} which is as yet
             insufficent to probe a significant part of the parameter
             space of SUGRA models. One needs more sensitive detectors
              with sensitvities 2-3 orders of magnitude better to
             sample a significant part of the parameter space of SUGRA
             models. (See in this context ref. \cite{cline}).
 
              As mentioned already the minimal supergravity grand
     unification gives a result for $\alpha_3(M_Z)$ which is about 2-3 std larger
              than the current world average \cite{chank,das}. However,
              because of the proximity of the GUT scale to the Planck scale
              one can expect corrections of size O($M_G/M_{Pl})$ where 
              $M_{Pl}$ is the Planck mass.
               For example, Planck scale corrections can modify the
               gauge kinetic energy function so that one has for the
               gauge kinetic energy  term -(1/4)$f_{\alpha\beta}
               F_{\alpha}^{\mu\nu}F_{\beta\mu\nu}$. 
               For the minimal SU(5) theory, Eq.(3) can give terms 
                $f_{\alpha\beta}=\delta_{\alpha\beta}+(c /2M_{Pl})
                d_{\alpha\beta\gamma}\Sigma^{\gamma}$ where
                $\Sigma$ is the the scalar field in the 24 plet of SU(5).
               After the spontaneous breaking of SU(5) and a re-diagonalization
               of the gauge kinetic energy function, one finds a splitting
               of the $SU(3)\times SU(2)\times U(1)$ gauge coupling
               constants at the GUT scale. These splittings can easily
               generate the 2 std correction to $\alpha_s(M_{Z})$.
                In fact, using the LEP data one can put constraints
               on c. One finds that \cite{das}$-1\leq c \leq 3$. 
          The Planck scale
               correction also helps relax the stringent constraint
               on $tan\beta$ imposed by $b-\tau$ unification. Thus in the
               absence of Planck scale correction one has that $b-\tau$
               unification requires $tan\beta$ to lie in two rather
               sharply defined corridors \cite{bbo}. One of these corresponds
               to a small value of $tan\beta$, i.e., $tan\beta \sim 2$
                 and the second a large value $tan\beta \sim 50$. This
                 stringent constraint is somewhat relaxed by the
                 inclusion of Planck scale corrections \cite{das}.

             SUSY grand unified models contain many sources of
             proton instability. Thus in addition to the
             p decay occuring via the exchange of superheavy  vector
             lepto-quarks,
             one has the possibility of p dacay from dimension (dim)
              4 (dim 3 in the superpotential) and dim 5
              (dim 4 in the superpotential) operators \cite{wein}.
              The lepto-quarks exchange would
              produce  $p\rightarrow e^+\pi^0$
              as its dominant mode with an expected lifetime \cite{gm} of
               $\sim 1\times
              10^{35\pm 1}{[M_X/10^{16}] }^{4}y $ where for SU(5)
              $M_X \cong 1.1 \times 10^{16}$
               while the sensitivity of Super Kamiokande (Super-K)
              to  this mode \cite{pdg,totsuka} is $1\times 10^{34}$.
              Thus the $e^+\pi^0$ mode may be at the edge of being accessible
              at Super-K.
                 Proton decay from dim 4 operators is much too rapid but is
                 easily forbidden by the imposition of R parity invariance.
                  The p decay from dim  5 operators is more involved.
                  It depends on
                  both the GUT physics as well as on the
                  low energy physics such as
                  the masses of the squarks and of the gauginos.
                  Analysis in supergravity unified models $^{63-65}$ shows that
                   one can make concrete predictions of the p decay
                   modes within these models. Thus one can show that
                  with the Higgs triplet mass $M_{H_3} < 10 M_G$,
                   the parameter space of minimal SU(5)
                  SUGRA will come close to being exhausted for
                  $m_0<1TeV$
                  and $m_{\tilde g}<1 TeV$ if Super-K \cite{totsuka}
                  and Icarus \cite{icarus} reach the
                  sensitivity for $\bar \nu K^+$ mode of 2$\times 10^{34}$
                  $y$ \cite{an94,oak}.

                   Precision determination of soft SUSY breaking parameters
                   can be utilized as a vehicle for the test of the
                  predictions of supergravity grand unification.
                  It was recently proposed that
                  precision measurement of the soft breaking parameters can also
                  act as a test of physics at the post GUT and string
                  scales \cite{planck}. Thus, for example,
                  if one has a concrete
                  model of the soft breaking parameters at the string scale
                  then these parameters can be evolved  down to the
                  grand unification scale leading to a predicted set of
                  non-universalities there. If the SUSY particle spectra and
                  their interactions are known with precision at the
                  electro-weak scale, then this data can be utilized
                  to test a specific  model at the post GUT or string
                  scales. Future colliders such as the LHC \cite{baer}
                  and the NLC \cite{tsukamoto}
                   will allow one to make
                  mass measurements with significant accuracy. Thus
                  accuracies  of up to a few percent
                  in the mass measurements will be possible at these 
                  colliders allowing
                  a test of post GUT and string physics up to an
                  accuracy of $\sim 10\%$ \cite{planck}.

\bigskip
\noindent   {\bf 5.~~ Conclusion}\\
\medskip

\indent
  Supergravity grand unification provides a framework for the
  supersymmetric unification of the electro-weak and the strong
  interactions where supersymmetry is broken spontaneously by
  a super Higgs effect in the hidden sector
   and the breaking communicated to the
  visible sector via gravitational interactions. The minimal version
  of the model based on a generation independent Kahler potential
  contains only four additional arbitrary parameters and one 
  sign in terms of which all the
  SUSY mass spectrum and all the SUSY interaction structure
   is determined. This model is thus very predictive. A brief summary
   of the  predictions and the phenomenological implications
   of the model was given. Many of the predictions of the model can be
   tested at current collider energies and at energies that would
   be achievable at future colliders. Other predictions of the
   model can be tested in non-accelerator experiments such as at
   Super Kamiokande and at Icarus, and by dark matter detectors. 
   We also discussed  here extensions of
   the minimal supergravity model to include non-unversalities in the
   soft SUSY breaking parameters. Some of the implications of these
   non-universalities on predictions of the model were discussed.
   Future experiments should be able  to see if  the predictions of
   supergravity unification are indeed verified in nature.

\bigskip

 \noindent
 {\bf Acknowledgements:} This work was supported in part by
 NSF grant numbers  PHY-9411543 and PHY-96020274.

 \end{document}